\documentclass{article}

\usepackage{arxiv}
\usepackage[utf8]{inputenc} 
\usepackage[T1]{fontenc}    
\usepackage{hyperref}       
\usepackage{url}            
\usepackage{booktabs}       
\usepackage{amsfonts}       
\usepackage{nicefrac}       
\usepackage{microtype}      
\usepackage{lipsum}
\usepackage[T1]{fontenc}
\usepackage{authblk}
\usepackage{graphicx}
\usepackage{float}
\usepackage{amsmath}
\usepackage{array}
\usepackage{multirow}
\usepackage{diagbox}
\usepackage{algorithm}
\usepackage{algorithmicx}
\usepackage{algpseudocode}
\usepackage[labelfont=bf]{caption}
\usepackage{csvsimple}
\usepackage[title]{appendix}

\usepackage{apacite}
\usepackage{natbib}
\usepackage{listings}
\usepackage{subcaption}
\usepackage{tablefootnote}
\AtBeginDocument{

}

\title{Understanding intra-day price formation process by agent-based financial market simulation: Calibrating the Extended Chiarella Model}
\author[1,2]{Kang Gao}
\author[2]{Perukrishnen Vytelingum}
\author[1]{Stephen Weston}
\author[1]{Wayne Luk}
\author[1]{Ce Guo}
\affil[1]{Department of Computing, Imperial College London\vspace{0.2cm}}
\affil[2]{Simudyne}

\begin{document}
\maketitle

\begin{abstract}
\hspace{0.3cm} This article presents XGB-Chiarella, a powerful new approach for deploying agent-based models to generate realistic intra-day artificial financial price data. This approach is based on agent-based models, calibrated by XGBoost machine learning surrogate. Following the Extended Chiarella model, three types of trading agents are introduced in this agent-based model: fundamental traders, momentum traders, and noise traders. In particular, XGB-Chiarella focuses on configuring the simulation to accurately reflect real market behaviours.  Instead of using the original Expectation-Maximisation algorithm for parameter estimation,  the agent-based Extended Chiarella model is calibrated using XGBoost machine learning surrogate. It is shown that the machine learning surrogate learned in the proposed method is an accurate proxy of the true agent-based market simulation. The proposed calibration method is superior to the original Expectation-Maximisation parameter estimation in terms of the distance between historical and simulated stylised facts. With the same underlying model, the proposed methodology is capable of generating realistic price time series in various stocks listed at three different exchanges, which indicates the universality of intra-day price formation process. For the time scale (minutes) chosen in this paper, one agent per category is shown to be sufficient to capture the intra-day price formation process. The proposed XGB-Chiarella approach provides insights that the price formation process is comprised of the interactions between momentum traders, fundamental traders, and noise traders. It can also be used to enhance risk management by practitioners.
\end{abstract}



\section{Introduction} \label{sectionintroduction}
\subsection{Motivation}

\hspace{0.3cm} In the past decade algorithmic trading has grown rapidly across the world and has become the dominant way securities are traded in financial markets, currently generating more than half of the volume of U.S. equity markets. Constantly improving computer technology and its application by both traders and exchanges, together with the evolution of market micro-structure, automation of price quotation and trade execution have together enabled faster trading. Consequently, intra-day price formation underpinning this trading process has become the focus of intense research attention in recent years as market participants attempt to gain greater insight into how prices are determined and hence improve trading performance.

\hspace{0.3cm} Price formation determines the price of an asset through interactions between buyers and sellers. It is at the core of the efficient and transparent operation of markets for goods and services. The balance between buyers and sellers provide an effective indicator of demand and supply in a market, where demand and supply are generally significant but not the only driving factors behind price movements. This is because the mechanisms of price discovery indicate what sellers are willing to accept and what buyers are willing to pay, so the price discovery process is concerned with finding an equilibrium (or near equilibrium) price that enables the greatest liquidity for that asset at a given point in time. Beyond supply and demand, attitudes to risk, volatilities, available information and market micro-structure all exert varying levels of influence on the price discovery process.

\hspace{0.3cm} \cite{lo2017adaptive} explains the process of price formation in properly functioning markets as generally involving market participants engaging in cause and effect reasoning along the lines of “if my strategy is x, then the market will respond with y, in which case I will respond with z…”. Even this simple process requires that the algorithms of buyers and sellers have some understanding of the other’s motives and incentives. Theoretically, this approach would imply that such chain reasoning could infinitely recurse. 

\hspace{0.3cm} However, \cite{sirignano2019universal} define a "price formation mechanism" as a high-level map representing the relationship between asset price and variables such as order flow and market price history. Modelling such a mechanism using stochastic differential equation models, machine learning prediction models and market micro-structure, all provide different ways to represent this map. However, an issue central to the price formation mechanism is the degree to which such a high-level map is universal. That is, whether the price formation mechanism is independent of the particular asset being considered. The universal existence of certain empirical stylized facts seems to be evidence supporting this universality traded. In this work, we present evidence for the existence of such a universal price formation mechanism by proposing the XGB-Chiarella method, which is able to reproduce realistic synthetic data for various stocks from different exchanges. The fact that XGB-Chiarella method is based on the same underlying agent-based model backs the existence of a universal price formation mechanism. 

\hspace{0.3cm} Our investigation of intra-day price formation process is through financial market simulation using agent-based models. Financial markets are obviously one of the most dynamic systems in existence. With huge potential academic and industrial value, financial market simulation in agent-based models is an exciting new field for exploring behaviours of financial markets. In an agent-based artificial financial market, heterogeneous agents (traders) trade a financial instrument through a realistic trading mechanism for price formation. Unlike traditional economic theories, there is no equilibrium assumption in agent-based financial markets. In addition, traders are no longer assumed to have rational behaviours as in traditional economic theories. The removal of these assumptions makes agent-based financial market simulation more realistic than traditional equilibrium-based economic and financial theories. Various agent-based simulators have been developed in literature. However, there are still gaps in creating ideal agent-based financial market simulators that are capable of generating realistic synthetic market data and shedding light on the intra-day price formation process. Specifically, most existing agent-based financial markets are of lower frequency such as daily or monthly. To investigate intra-day price formation process, higher simulation frequency is needed. 

\hspace{0.3cm} To ensure realism in the generated intra-day financial time series data, parameters of the agent-based model must be calibrated to be as close to real market data as possible. Realistic simulated market data are supposed to exhibit certain characteristics known as stylised facts, which are universally observed in historical financial market data. Some stylised facts originate from behaviours of market participants, while others could be natural consequences of market structure design. Examples of stylised facts include fat tails of returns, volatility clustering, etc. Parameter calibration with respect to certain stylised facts can be extremely challenging due to the huge parameter space and complexity in designing explicit optimization objective function. 

\hspace{0.3cm} To sum up, there are still two challenges of great interest in this field:
\begin{itemize}
\item C1: To implement an agent-based financial market simulator which allows for the investigation of intra-day price formation process in financial market.
\item C2: To calibrate the agent-based financial market simulator to ensure realism and reproduce common stylised facts.
\end{itemize}

To address the two challenges, we developed XGB-Chiarella, which is a novel approach to developing and calibrating an intra-day financial market simulator to narrow the existing gaps. The XGB-Chiarella methodology has two essential components: the underlying mathematical model for simulation and the calibration workflow with surrogate modelling. For the underlying model, the simple but powerful Extended Chiarella model (\citealp{MAJEWSKI2020103791}), which consists of fundamental traders, momentum traders and noise traders, is used as the underlying model for the XGB-Chiarella method. For the calibration workflow, the method utilises XGBoost\footnote{XGBoost (\citealp{chen2016xgboost}) is a highly flexible and efficient machine learning algorithm based on an ensemble of decision trees.} algorithm to build a machine learning surrogate for the purpose of model calibration. We will show that even with only one agent for each type of trader, the XGB-Chiarella method is able to generate realistic price series simulations after proper model parameter calibration process.

\subsection{Background and Related Work}
\hspace{0.3cm} With the rapid development of modern financial markets, price formation process in financial market has been of great interest to both researchers and practitioners for many years. One group of price formation process literature is based on the equilibrium state of financial market. \cite{faias2011equilibrium} propose a pure exchange economy with a finite number of types of agents and commodities. They analyse the equilibrium price formation in a differential information market, where traders have incomplete and asymmetric information. \cite{jackson1991equilibrium} shows that it is possible to have an equilibrium with fully revealing prices and costly information acquisition if the price formation process is modelled explicitly and traders are not price-takers. Price formation process is also extensively investigated from the perspective of double auction market. \cite{cason1996price} present 14 laboratory experiments that examine the price formation process in the continuous double auction. It is shown that participants in their double auction market experiments succeed in discovering prices that would achieve most of the exchange surplus. The same is true even with no auctioneer and with traders possessing various private information.  \cite{gerety1994price} examine the relationships between market structure and stock price formation. It is shown that trading mechanism and price formation provide different explanations for the greater volatility at the opening of trading.

\hspace{0.3cm} Unlike the above works, in this article we analyse the price formation process in financial markets from another aspect, which focuses on the so-called trend and value effects. Trend and value effects are indispensable when it comes to price formation process in financial market. The two interactive effects pervade all financial markets. Trend refers to the price behaviour that positive (negative) returns are more likely to be followed by positive (negative) returns. Value means that the asset price will converge to the intrinsic value of the asset, which indicates that assets with prices higher (lower) than their fundamental value tend to achieve negative (positive) future returns. The trend and value effects correspond to two types of market participants: fundamental trader and momentum trader. Fundamental trader reacts to the difference between fundamental value and market price, while momentum trader reacts to price trends. Lots of models investigating trend and value effects are analysed in the literature. 

\hspace{0.3cm} \cite{beja1980dynamic} build a model which proves that the so-called "speculation on the price trend" plays an important role in the formation of dynamic price behaviours. It is shown that speculative trading can accelerate the convergence to the equilibrium state, but it can also lead to price oscillations and market instability. \cite{frankel1986understanding} present a model containing fundamentalist and chartist to explain the dollar price in the early 1980s. Not constrained by the assumption of utterly rational behaviours, in this model each type of trader performs the specific task in a reasonable and realistic way. Their model provides a framework for explaining price formation process in a variety of asset markets. \cite{ZEEMAN197439} shows that the unstable behaviours in various financial markets can be credited to the interactions between fundamental traders and momentum traders. In one famous paper, \cite{chiarella1992dynamics} proposes the so-called Chiarella model which consists of fundamentalists and chartists in the artificial financial market. It is shown that the Chiarella model is capable of generating a number of dynamic regimes in financial market which are consistent with empirical evidence. Based on the Chiarella model, \cite{MAJEWSKI2020103791} propose an Extended Chiarella model by adding noise traders and changing the demand function of fundamentalists. The Extended Chiarella model investigates the co-existence and interaction between trend and value effects in the framework of agent-based models. The model parameters are estimated using an Expectation-Maximisation algorithm. Nevertheless, all the above models share some common drawbacks. Firstly, all the above models are estimated by mathematical derivation. Consequently, all those models generate theoretical results instead of actual simulated results. Secondly, existing models are used to explain daily or monthly price behaviours. There still exists a large gap in successfully explaining intra-day price formation process in terms of trend and value effects.

\hspace{0.3cm} In this work, we examine the intra-day price formation process in the framework of agent-based models, where trend and value effects are represented by heterogeneous traders. Given an agent-based model, how to effectively calibrate the model to real data is still an open challenge. Successful calibration of an agent-based model enables the model to generate qualitative or quantitative properties that are observed consistently in empirically-measured data and cannot be reproduced using traditional equilibrium based approaches (\citealp{lebaron2006agent}). Lots of calibration methods have been proposed in the literature. For some simple models such as the CATS model in (\citealp{bianchi2007validating}), model parameters can be read directly from the data.  Analytical method is another class of agent-based model calibration method. For example, \cite{MAJEWSKI2020103791} apply Expectation-Maximisation method to get the maximum likelihood estimation of the parameters of the Extended Chiarella model. However, for most complex models, model parameters are not directly observable. In addition, most agent-based models that are of interest are incompatible with calibration methods that require analytical solutions. In those cases the only choice for model calibration is simulation-based method. The most common simulation-based calibration method is the simulated minimum distance method and its variations (\citealp{GRAZZINI2015148}). The simulated minimum distance method involves the construction of an objective function that measures the distance between real data and simulated data for a given set of model parameters. Optimisation methods are subsequently applied to minimise the distance to get an optimal set of model parameters. In the context of economic agent-based models, popular distance measures include weighted sums of the squared errors between empirical moments and simulated moments. \cite{FRANKE2009804} applies the method of simulated moments to estimate the parameters of an agent-based asset pricing model. Their moment selection emphasizes the reflection of certain stylised facts in financial markets, such as the fat tails and autocorrelation patterns of the daily returns in stock price time series. To explore structural stochastic volatility, \cite{FRANKE20121193} employ the method of simulated moments to estimate parameters of different candidate agent-based asset pricing models. They also take into consideration the proportion of Monte Carlo simulation runs that yield moments within the empirical moments confidence intervals.

\hspace{0.3cm} Another large obstacle in the development of robust and widely-applicable agent-based model calibration strategies is the computational complexity. Most agent-based models of interest are computational costly to simulate and the situation is even worse when it comes to large-scale agent-based models. One possible solution to address this challenge is the use of surrogate modelling to help guide parameter space exploration and thus avoid a large amount of intensive agent-based model simulations. Examples of surrogate modelling include kriging (\citealp{rasmussen2003gaussian}) and machine learning surrogate approach (\citealp{LAMPERTI2018366}). \cite{dosi2018robustness} apply kriging method to enable a global exploration of the parameters space for a multi-firm evolutionary simulation model. Sensitivity analysis is also carried out in this kriging framework. \cite{LAMPERTI2018366} build machine learning surrogate models to approximate two agent-based models. Experimental results show that their XGBoost-based machine learning surrogate achieves high accuracy in approximating the relationship between model parameters and model outputs. The approach of machine learning surrogate modelling is also the foundation of the calibration method in this article.

\subsection{Our Contributions}
\hspace{0.3cm} In this paper, we propose XGB-Chiarella - a method for developing and calibrating an agent-based market simulator to generate realistic intra-day financial market data. The underlying model is the Extended Chiarella model (\citealp{MAJEWSKI2020103791}). We adapted the machine learning surrogate modelling approach in \cite{LAMPERTI2018366} to calibrate the agent-based financial market simulator. By reproducing realistic synthetic data for various stocks, we show that there is a universal intra-day price formation process which involves trend and value effects. The main contributions are:
\begin{itemize}
\item Addressing challenge C1: The Extended Chiarella model was originally tested on monthly data. To address Challenge C1, we further tested the Extended Chiarella model in intra-day minute data to explain the intra-day price formation process. The model is tested extensively in more than 75 stocks from three major exchanges in the world: Nasdaq, London Stock Exchange, and Hong Kong Stock Exchange. For all stocks of different exchanges, the experimental results are similar and the simulator can all produce realistic simulated financial time series. This shows that trend and value effects exist universally in the stock market price formation process, regardless of the exchanges.
\item Addressing challenge C2: Instead of the Expectation-Maximisation algorithm, we propose a novel application of a recent machine learning surrogate modelling approach (\citealp{LAMPERTI2018366}) to calibrate the Extended Chiarella model, which addresses challenge C2. The foundation for this calibration method is the surrogate modelling approach in \cite{LAMPERTI2018366}. Instead of building an XGBoost classifier to predict positive calibration and negative calibration, we train an XGBoost regressor to innovatively predict the actual distance between simulated stylised facts and historical stylised facts. The distance involves not only the return distribution but also the autocorrelations between returns and squared returns. In addition, exploration-exploitation mechanism is introduced in the iterative process of selecting new points in parameter space. In terms of stylised facts distance, results show that the proposed method performs much better than the baseline Expectation-Maximisation estimation algorithm. 
\end{itemize}

\section{Model Architecture}
\hspace{0.3cm} This section presents the set-up and components of the agent-based financial market simulator. 

\subsection{Model Set-up} \label{modelsetup}
\hspace{0.3cm} We denote the price of a stock at time $t$ as $P_t$. The total signed volume traded on the market from $t$ to $t+\Delta$ constitutes the cumulative demand imbalance in the same period. This quantity is denoted as  $D(t, t+\Delta)$. This aggregated demand depends on the trading strategies of various types of market participants. Following \cite{MAJEWSKI2020103791} and \cite{kyle1985continuous}, the price dynamics is assumed to be governed by a linear price impact mechanism:
\begin{equation} \label{kyleslambda}
P_{t+\Delta} - P_t = \lambda D(t, t+\Delta)
\end{equation}
where $\lambda$ is called "Kyle's lambda", which is related to the liquidity of the market and is a first-order approximation of market price sensitivity to market demand and supply. The market participants are assumed to be heterogeneous in their trading decisions. Following the Extended Chiarella model (\citealp{MAJEWSKI2020103791}), we populate our model with fundamental traders, momentum traders, and noise traders. Since traders of the same type exhibit same behaviours, we only use one agent for each type of trader. This single agent represents the corresponding type of traders in the market. With only three agents in the model, simulation process is significantly accelerated.  Each trader is associated with some parameters that control the trading behaviours and the amount of demand generated by the corresponding trader group. We will show the calibration of these parameters in later sections.


\subsection{Fundamental Trader}
\hspace{0.3cm} Fundamental traders make their trading decisions based on the perceived fundamental value of the stock. The fundamental value is denoted as $V_t$. A fundamentalist will tend to buy a stock if the stock is under-priced ($V_t - P_t > 0$), otherwise it will tend to sell the stock. Following the convention in \cite{chiarella1992dynamics}, in this work we assume the aggregated demand of fundamental traders is proportional to the level of mispricing. In other words, the aggregated demand of fundamentalists is $\kappa(V_t - P_t)$, where $\kappa$ controls the overall demand generated by fundamental traders. The fundamental value $V_t$ is an exogenous signal that is input to the model. 

\subsection{Momentum Trader}
\hspace{0.3cm} Momentum traders are also called "Chartists". This group of traders buy and sell financial assets after being influenced by recent price trends. The assumption is to take advantage of upward or downward trend of the stock prices until the trend starts to fade. Instead of looking at the fundamental value of the stock, momentum traders focus more on recent price action and price movement. If the stock price has been recently rising, a long position is established; otherwise momentum traders will enter a short position.

\hspace{0.3cm} There are lots of methods to estimate the momentum of stock prices. A common trend signal is the exponentially weighted moving average of past returns with decay rate $\alpha$. This trend signal is denoted by $M_t$:
\begin{equation} \label{momentumequation}
M_{t} = (1 - \alpha)M_{t-1} + \alpha(p_t - p_{t-1})
\end{equation}
where $\alpha$ is the decay rate. Given the trend signal $M_t$, the demand function of momentum traders is denoted as $f(M_t)$. The demand function $f(M_t)$ must satisfy two conditions:
\begin{itemize}
\item $f(M_t)$ is increasing.
\item $f''(M_t) * M_t < 0$
\end{itemize}
where the first condition is consistent with the nature of momentum trading and the second condition imposes the risk-averse assumption to momentum traders. Consistent with \cite{chiarella1992dynamics}, here we choose $f(M_t) = \beta \tanh(\gamma M_t)$ with the requirement that $\gamma > 0$. $\gamma$ represents the saturation of momentum traders' demand when momentum signals are very large. This phenomenon is partly due to for example budget constraints and risk aversion, which is prevalent in real chartists. $\beta$ controls the overall demand generated by momentum traders. $\beta$ is also assumed to be positive, i.e. the demand of momentum traders is positive when the momentum signal ($M_t$) is positive, otherwise the demand is negative. The choice of this demand function for momentum traders strictly satisfies the two requirements. 

\subsection{Noise Trader}
\hspace{0.3cm} Another group of market participants is noise traders. They are designed so as to capture other market activities that are not reflected by trend-following and value investing. As a result, the cumulative demand of noise traders can be described by a random walk. The random walk is also multiplied by a parameter $\sigma_N$, which controls the overall demand level from noise traders. Mathematically, for each step the demand from noise traders is sampled from a normal distribution with zero mean and standard deviation $\sigma_N$.

\subsection{Model Dynamics}

\subsubsection{Simulation Process}
\hspace{0.3cm} Using mathematical formulas to approximate the supply and demand generated by all the participating traders, the resulting model dynamics for $\Delta t \rightarrow 0$ can be described by the following dynamical system. Note that here the fundamental value $V_t$ is an exogenous signal that is input to the model, which is a major adaption from the original Extended Chiarella model. In addition, there are only three agents in our simulation, namely fundamental trader, momentum trader, and noise trader. Each trader creates a demand corresponding to one term in Equation~(\ref{demandmodel}). Our simulation results will show that this model is able to generate very realistic artificial financial price time series, even though there are only three agents included in the model.

\begin{equation} \label{demandmodel}
\begin{aligned}
D(t, t+\Delta t) &=  \underbrace{\kappa^\prime(V_t - P_t)\Delta t}_{Fundamental} + \underbrace{ \beta^\prime\tanh(\gamma M_t)\Delta t}_{Momentum} + \underbrace{ \sigma_{N}^\prime \epsilon_t \sqrt{\Delta t}}_{Noise} \\
dM_t &= -\alpha M_t\Delta t + \alpha dP_t\\
\end{aligned}
\end{equation}

where $\epsilon_t$ follows standard normal distribution. Substitute Equation~(\ref{kyleslambda}) into the above equations, we can get:

\begin{equation} \label{pricemodel}
\begin{aligned}
dP_t &=  \kappa(V_t - P_t)\Delta t + \beta\tanh(\gamma M_t)\Delta t + \sigma_{N} \epsilon_t \sqrt{\Delta t}\\
dM_t &= -\alpha M_t\Delta t + \alpha dP_t\\
\end{aligned}
\end{equation}

where $\kappa$, $\beta$, $\sigma_N$  equal to $\lambda\kappa^\prime$, $\lambda\beta^\prime$, $\lambda\sigma_{N}^\prime$, respectively. Furthermore, the simulator runs according to a discrete-time version of model~(\ref{pricemodel}), in which $\Delta t$ is 1, corresponding to 1 unit of the smallest simulation time interval:

\begin{equation} \label{discretemodel}
\begin{aligned}
P_t - P_{t-1} &=  \kappa(V_t - P_t) + \beta\tanh(\gamma M_t) + \sigma_{N} \epsilon_t \\
M_t &= (1 - \alpha) M_{t-1} + \alpha (P_t - P_{t-1})\\
\end{aligned}
\end{equation}

\hspace{0.3cm} The whole simulation runs according to Equation~(\ref{discretemodel}). For each step, each trader collects and processes market information. Internal variables associated with each trader are calculated. According to agent types and values of internal variables, demands are generated by the traders. The price evolves according to Equation~(\ref{kyleslambda}).

\subsubsection{Fundamental Value from Kalman Smoother}
\hspace{0.3cm} The only remaining unknown variable is the fundamental value of the stock. According to Equation~(\ref{discretemodel}), the simulation can proceed only if fundamental value is known and is exogenously input to the model. One difficulty is the non-observability of the fundamental value. According to the economic literature, the fundamental value of a stock equals to the expected value of discounted dividends that the company will pay to the shareholders in the future. However, this methodology requires extremely strong assumptions on the future dynamics of the stock dividends. Furthermore, this approach can never reflect the intra-day change of fundamental value, while the consensus fundamental value can indeed vary during the trading day due to the continuous feed of events and news. 

\hspace{0.3cm} In this paper, we propose a novel method which is to apply Kalman Smoother (\citealp{1556088}) directly to the stock price time series to get the hidden fundamental value. Note that Equation~(\ref{discretemodel}) is a linear dynamical system in $V_t$, which is treated as a hidden variable of the system. Here the observations are the actual prices traded in real market. The specific Kalman Smoothing algorithm used here can be found in \cite{byron2004derivation}. The algorithm is already implemented in Python package "pykalman". 

\section{Model Calibration}
\hspace{0.3cm} In this section we present the methodology for calibrating the agent-based financial market simulator. Calibration means finding an optimal set of model parameters to make the model generate most realistic simulated financial market. Firstly we describe the real data and the associated stylised facts in financial markets. Next we define the distance between historical and simulated stylised facts, which acts as the loss function in the calibration process. Finally the machine learning surrogate modelling for parameter space exploration is presented in detail.

\subsection{Data} \label{data}
\hspace{0.3cm} In the model calibration process, real financial market data is essential to set up the calibration target. We collected stock price data of 75 stocks from three major exchanges in the world - Nasdaq, London Stock Exchange, and Hong Kong Stock Exchange. Our dataset comprises intra-day minute price data for those 75 stocks, spanning the entire trading period from September 20th, 2021 to November 30th, 2021. Detailed stocks from each exchange are shown in Table~\ref{stockcode}.

\begin{table}[H]
\centering
\caption{Specific stocks from three exchanges in the dataset.}
\begin{tabular}{ | m{2cm}<\centering | m{14cm}<\centering |} 
\hline
Exchange &  Stocks\\
\hline
Nasdaq &   AAPL, HUT, AMD, AAL, MSFT, INTC, UBER, NVDA, SOFI, DKNG, WISH, HON, FB, TSLA, MRNA, AFRM, LCID, CMCSA, HBAN, TLRY, MU, CSX, CSCO, JD, UAL\\ 
\hline
LSEG &  LLOY.L, VOD.L, RR.L, GLEN.L, IAG.L, HSBA.L, BP.L, PRU.L, DGE.L, LGEN.L, AAL.L, AV.L, RDSA.L, STAN.L, ULVR.L, BHP.L, BATS.L, RIO.L, GSK.L, NG.L, REL.L, EZJ.L, JET.L, SMT.L, BRBY.L \\ 
\hline
HKEX &  0700.HK, 3690.HK, 9988.HK, 1299.HK, 1211.HK, 2020.HK, 2269.HK, 0175.HK, 2331.HK, 1024.HK, 0916.HK, 2318.HK, 1918.HK, 9618.HK, 9999.HK, 2333.HK, 9888.HK, 0836.HK, 1919.HK, 0388.HK, 3968.HK, 1772.HK, 1171.HK, 0005.HK, 0027.HK\\ 
\hline
\end{tabular}
\label{stockcode}
\end{table}

\subsection{Stylised Facts and Loss Function}
\hspace{0.3cm} Financial price time series data display some interesting statistical characteristics that are commonly called stylised facts. According to \cite{sewell2011characterization}, stylized facts refer to empirical findings that are so consistent (for example, across a wide range of financial instruments and different time periods) that they are accepted as truth. A stylized fact is a simplified presentation of an empirical finding in financial market. A successful and realistic financial market simulation is capable of reproducing various stylised facts. These stylised facts include fat-tailed distribution of returns, autocorrelation of returns, and volatility clustering. The loss function used in the calibration process is constructed by measuring the distance between historical and simulated stylised facts.

\subsubsection{Fat-tailed distribution of returns}
\begin{figure}[H]
\centering
\begin{minipage}[t]{0.48\textwidth}
\centering
\includegraphics[width=7cm, height=5cm]{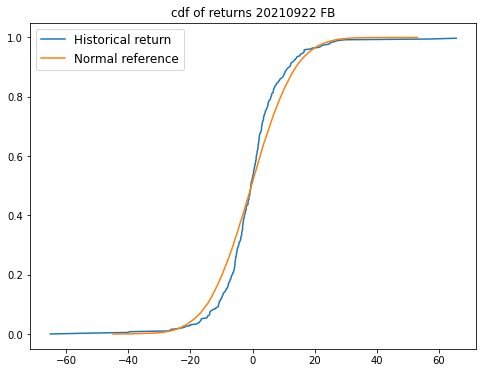}
\end{minipage}
\begin{minipage}[t]{0.48\textwidth}
\centering
\includegraphics[width=7cm, height=5cm]{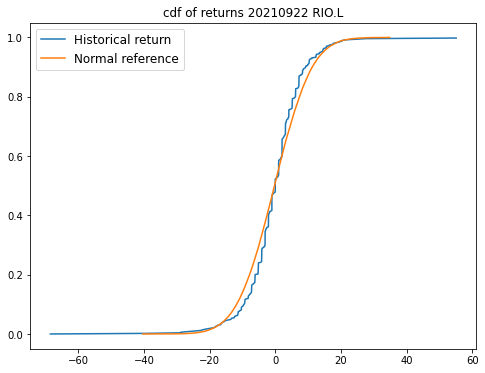}
\end{minipage}
\caption{Cumulative distribution function of historical returns in two stocks (FB, RIO.L) for one trading day, in comparison with a normal distribution reference.}
\label{kurtosisgraph}
\end{figure}

\hspace{0.3cm} The distributions of price returns have been found to be fat-tailed across all timescales. In other words, the return distributions exhibit positive excess kurtosis. Understanding positively kurtotic return distributions is important for risk management, since large price movements are much more likely to occur than in commonly assumed normal distributions. In this paper we focus on intra-day minute price returns. Excess kurtosis of the distribution of minute-level returns is calculated for each stock at each trading day. Table~\ref{returnkurtosis} presents the average excess kurtosis for the intra-day minute-level return distributions of six randomly chosen stocks. For each stock, the return distribution has significantly positive excess kurtosis, proving the fat-tail characteristic of return distributions. Figure~\ref{kurtosisgraph} also shows a comparison between cumulative distribution function of historical returns in two stocks for one trading day and that of a normal distribution with identical mean and standard deviation. Similar results are found in all other stocks from the three exchanges.

\begin{table}[H]
\centering
\caption{Average excess kurtosis for return distributions of five stocks.}
\begin{tabular}{ | m{3cm}<\centering | m{1.5cm} | m{1.5cm} | m{1.5cm} | m{1.5cm} | m{1.5cm} | m{1.5cm} |} 
\hline
Stock & FB  & AAPL & VOD.L & JET.L & 9999.HK & 0700.HK \\
\hline
Excess Kurtosis & 5.51  & 5.64 & 16.20 & 9.60 & 9.49 & 4.66 \\ 
\hline
\end{tabular}
\label{returnkurtosis}
\end{table}

\subsubsection{Autocorrelation of returns}
\begin{figure}[H]
\centering
\includegraphics[width=16.5cm, height=5cm, angle=0]{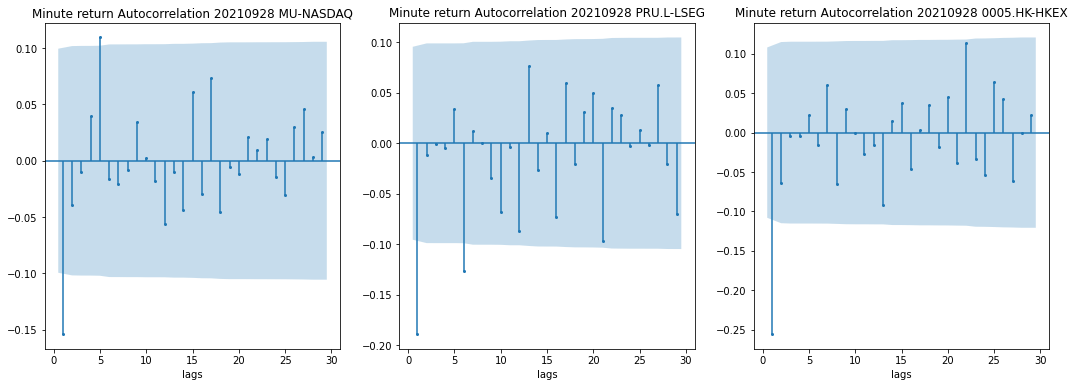}
\caption{Autocorrelation patterns for returns of three stocks from the three exchanges.}
\label{acf1graph}
\end{figure}

\hspace{0.3cm} Autocorrelation is defined to be a mathematical representation of the degree of similarity between a time series and a lagged version of the same time series. It measures the relationship between a variable's past values and its current value. Take first-order autocorrelation for example. A positive first-order autocorrelation of returns indicates that a positive (negative) return in one period is prone to be followed by a positive (negative) return in the subsequent period. Instead, if the first-order autocorrelation of returns is negative, a positive (negative) return will usually be followed by a negative (positive) return in the next period. It is observed that the return series lack significant autocorrelation, except for weak, negative autocorrelation on very short timescales. \cite{mcgroarty2019high} show that the negative autocorrelation of returns is significantly stronger at a smaller time horizon and disappears at a longer time horizon. Examination of our data supports this stylised fact. Figure~\ref{acf1graph} shows the autocorrelation function of minute-level return time series for several stocks up to lag 30. We can see that the autocorrelation is significantly negative for small lags, and the negative autocorrelation gradually disappears for larger lags.

\subsubsection{Volatility clustering}
\begin{figure}[H]
\centering
\includegraphics[width=16.5cm, height=5cm, angle=0]{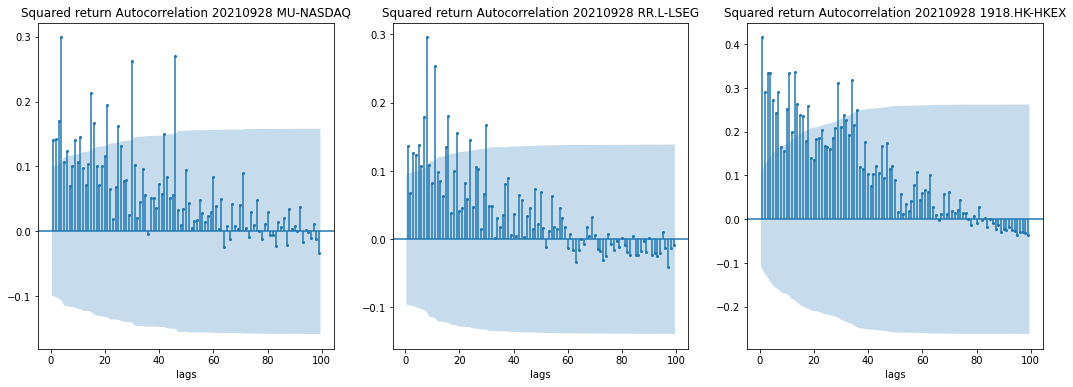}
\caption{Autocorrelation patterns for squared returns of three stocks from the three exchanges.}
\label{acf2graph}
\end{figure}

\hspace{0.3cm} Financial price returns often exhibit the volatility clustering property: large changes in prices tend to be followed by large changes, while small changes in prices tend to be followed by small changes. This property results in persistence of the amplitudes of price changes (\citealp{cont2007volatility}). It is found that the volatility clustering property exists on timescales varying from minutes to weeks and months. Volatility clustering also refers to the long memory of square price returns (\citealp{mcgroarty2019high}). Consequently, volatility clustering can be manifested by the slow decaying pattern in the autocorrelations of squared returns. Specifically, for short lags the autocorrelation function of squared returns is significantly positive, and the autocorrelation slowly decays with the lags increasing. Figure~\ref{acf2graph} shows the autocorrelation patterns for squared returns of several randomly chosen stocks from the three exchanges. Autocorrelation for squared returns of other stocks also exhibit similar patterns. It is shown that the volatility clustering stylised fact exists universally in financial markets. 

\subsubsection{Stylised Facts Distance as Loss Function}
\hspace{0.3cm} The target for agent-based model calibration is to find an optimal set of model parameters to make the model generate realistic simulated financial market. To solve this optimization problem, it is essential to have a metric that is able to quantify the "realism" of a simulated financial market. First of all, a realistic simulated financial market must exhibit similar characteristics to real financial market, such as the return distribution and volatility level. In addition, realistic simulated financial data are also required to reproduce other stylised facts such as the autocorrelation patterns in returns and squared returns. Here we design a stylised facts distance to quantify the similarities between simulated and historical financial data. The stylised facts distance is the weighted sum of four quantities: Kolmogorov–Smirnov statistic of return distributions, volatility difference, autocorrelation difference of returns and autocorrelation difference of squared returns:

\begin{equation} \label{rawdistance}
\begin{aligned}
D &= w_1 * KS + w_2 * \Delta_{V} + w_3 * \Delta_{ACF^1} + w_4 * \Delta_{ACF^2} \\
\end{aligned}
\end{equation}

Detailed calculations of the four quantities in the stylised facts distance are introduced below.

\hspace{0.3cm} Kolmogorov–Smirnov statistic is a quantity from Kolmogorov–Smirnov test in statistics. Kolmogorov–Smirnov test is a non-parametric test of the equality of probability distributions that can be used to compare two samples. Here the two samples are simulated returns and historical returns, respectively. The Kolmogorov–Smirnov statistic quantifies a distance between the distribution functions of simulated returns and historical returns. Recall that the empirical distribution function is an estimate of the cumulative distribution function that generated the points in the sample. Let $F_s(x)$ and $F_h(x)$ denote the empirical distribution functions of simulated returns and historical returns, respectively. Then the Kolmogorov–Smirnov statistic is calculated as follows:

\begin{equation} \label{ksstatistics}
\begin{aligned}
KS &=  \sup_{x} |F_s(x) - F_h(x)|\\
\end{aligned}
\end{equation}

where $\sup_{x}$ is the supremum of the set of differences. Intuitively, the statistic represents the largest absolute difference between the two distribution functions across all x values. The smaller the statistic is, the more similar the simulated and historical returns are. The inclusion of Kolmogorov–Smirnov statistic in the stylised facts distance addresses the requirements that the simulated data have similar return distribution to historical data and exhibit fat tails of returns.

\hspace{0.3cm} The second part of the stylised facts distance is the absolute volatility difference between simulated returns and historical returns:
\begin{equation} \label{voldiff}
\begin{aligned}
\Delta_{V} &=  |V_s - V_h| \\
\end{aligned}
\end{equation}
where $V_s$ and $V_h$ denote simulated volatility and historical volatility, respectively. This part addresses the requirement that simulated financial market should be similar to real market in terms of volatility.

\hspace{0.3cm} The third part of the stylised facts distance is the difference between simulated and historical autocorrelations of returns. This part in the stylised distance measures the model's ability of reproducing autocorrelation patterns commonly found in historical returns. It is shown that financial price return time series lack significant autocorrelation, except for short time scales, where significantly negative autocorrelations exist. This phenomenon is backed by our historical minute-level returns data. For very small lags the autocorrelations are negative, while for larger lags the autocorrelations become insignificant. To measure the distance in autocorrelation patterns between simulated data and historical data, we invoke the autocorrelation function of returns and calculate the average absolute difference between autocorrelations of simulated return time series and historical return time series for various lags:
\begin{equation} \label{acf1diff}
\begin{aligned}
\Delta_{ACF^1} &=  \frac{ \sum\limits_{l \: in \: lags} |ACF_s(l, r) - ACF_h(l, r)| } { |lags|} \\
\end{aligned}
\end{equation}

where $ACF_s(l, r)$, $ACF_h(l, r)$ are the autocorrelation function of lag $l$ for simulated returns and historical returns, respectively. $|lags|$ denotes the number of lags used in the calculation. Because the empirical autocorrelations are negative for very small lags and close to zero for larger lags, it is not necessary to consider all of the autocorrelation coefficients. Empirical evidence suggests that the autocorrelation pattern is well represented by the coefficients for three lags: 1, 10, 20. Also, to reduce the effects of accidental outliers, the autocorrelation function is smoothed by calculating the three-lag average. That is, the lag 1 autocorrelation is calculated as the average autocorrelation of lag 1, 2, 3, and so as the calculation for lag 10 and lag 20. In total, autocorrelations of 9 lags (1, 2, 3, 10, 11, 12, 20, 21, 22) are considered and included in the calculation.

\hspace{0.3cm} The last part of the stylised facts distance is the difference between simulated and historical autocorrelations of squared returns. The replication of autocorrelation patterns in squared returns indicates the model's capability to reproduce the volatility clustering stylised fact. It is shown empirically that large price changes tend to be followed by other large price changes, known as the volatility clustering phenomenon. Consequently, though there is generally no significant patterns in autocorrelations of returns, the autocorrelations of squared returns are significantly positive, especially for small time lags. Also, as time lag increases, the autocorrelation of squared returns displays a slowly decaying pattern, as shown in Figure~\ref{acf2graph}. Similar to the difference between autocorrelations of returns $\Delta_{ACF^1}$, the difference between autocorrelations of squared returns is calculated as follows:

\begin{equation} \label{acf2diff}
\begin{aligned}
\Delta_{ACF^2} &=  \frac{ \sum\limits_{l \: in \: lags} |ACF_s(l, r^2) - ACF_h(l, r^2)| } { |lags|} \\
\end{aligned}
\end{equation}

where $ACF_s(l, r^2)$, $ACF_h(l, r^2)$ are the autocorrelation function of lag $l$ for simulated squared returns and historical squared returns, respectively. $|lags|$ denotes the number of lags used in the calculation. Unlike the case for autocorrelations of returns calculation, here we use a different $lags$. Because empirical autocorrelations of squared returns are significantly positive and slowly decaying, we consider the autocorrelations of squared returns from a minimal lag length of 1 minute up to a maximal lag length of 20 minutes. In total, the autocorrelations of 20 lags (1, 2, 3, ..., 18, 19, 20) are considered and included in the calculation.

\hspace{0.3cm} The above four parts, along with the corresponding weights, constitute the stylised facts distance in Equation~(\ref{rawdistance}). In our experiments, there is no preference for any one of the four stylised facts. Thus all the weights are equal to 1. Since it happens that the four quantities are of the same orders of magnitude, there is no need to adjust weights either. Also note that the stylised facts distance is a function of model parameters. In other words, given a set of model parameters, there is a unique stylised facts distance calculated from the simulated time series, which correspond to that particular set of model parameters. Let $\theta$ denote the vector of model parameters to be estimated, Equation~(\ref{rawdistance}) can be rewritten as:
\begin{equation} \label{distancefunction}
\begin{aligned}
D(\theta) &= w_1 * KS(\theta) + w_2 * \Delta_{V}(\theta) + w_3 * \Delta_{ACF^1}(\theta) + w_4 * \Delta_{ACF^2}(\theta) \\
\end{aligned}
\end{equation}
The smaller $D(\theta)$ is, the more realistic the simulation is. Thus $D(\theta)$ serves as the loss function that the calibration method aims to minimize by finding an optimal set of model parameters. Let $\Theta$ denote the admissible set for model parameter vector $\theta$, the calibration target is to find the optimal model parameter vector $\hat{\theta}$ that minimizes the stylised facts distance:
\begin{equation} \label{thetahat}
\begin{aligned}
\hat{\theta} &= \arg \; \min_{\theta \in \Theta} \; D(\theta) \\
\end{aligned}
\end{equation}
The calibration method is presented in detail in subsequent sections.

\subsection{Surrogate Modelling Calibration Method}
\subsubsection{Specification}
\hspace{0.3cm} We propose a novel surrogate modelling approach to calibrate the model parameters. That is, to find an optimal set of model parameters to minimize the loss function - stylised facts distance. The approach is adapted from the machine learning surrogate modelling methodology in \cite{LAMPERTI2018366}. The original methodology mainly trains an XGBoost classifier to classify a positive calibration or negative calibration (\citealp{LAMPERTI2018366}), while in our approach we train an XGBoost regressor to directly approximate the mapping from model parameters to stylised facts distance. Another novelty in our approach is the introduction of exploration-exploitation mechanism in selecting new points in parameter space. The advantage of surrogate modelling approach is that it significantly reduces the computational cost of calibrating an ABM with several model parameters. Using only a limited budget (N) of ABM evaluations, the XGBoost surrogate model is proved to be a fairly good approximation of the mapping from model parameters to the target stylised facts distance. The surrogate provides a costless way to predict the model's response and allows for efficiently finding the optimal point in the model parameter space that minimizes the loss function.

\hspace{0.3cm} With regard to the Extended Chiarella simulation model, during calibration the model can be represented as a mapping M: $\theta \rightarrow D(\theta)$ from a vector of model parameters $\theta$ into the stylised facts distance $D(\theta)$. Generally the number of parameters ranges from several to dozens. In our case, we have three parameters to calibrate: $\kappa$, $\beta$, and $\sigma_N$. The values for parameter $\gamma$ and $\alpha$ are fixed in advance. Parameter $\alpha$ indicates the typical horizon of trend computation for momentum traders. Following the original Extended Chiarella model (\citealp{MAJEWSKI2020103791}), $\alpha$ is relatively small to represent a relatively low frequency momentum trader. Here we take the value 0.1 for $\alpha$. Other values of $\alpha$ are also tested, and similar results are obtained as long as $\alpha$ is smaller than 0.4. For larger $\alpha$ value, the low frequency momentum traders will become high frequency momentum traders, which would be an extension of the Extended Chiarella model. We will present this extension in a separate paper. For $\gamma$, the value is fixed to be 10. Since $\gamma$ appears in the same term as $\beta$ in Equation~(\ref{discretemodel}), the model calibration will have difficulties pinning down the value of parameters $\gamma$ and $\beta$ if both parameters are to be estimated. As a result, fixing $\gamma$ significantly improves the robustness of the model calibration process. We have also tested other values of $\gamma$ and the results are similar. 

\hspace{0.3cm} Instead of classifying a positive calibration or negative calibration as in the original method (\citealp{LAMPERTI2018366}), here our calibration objective is obvious: finding an optimal set of values for $\kappa$, $\beta$, and $\sigma_N$ to minimize the stylised facts distance. Our calibration measure is a real-valued number providing a quantitative assessment of the realism of the simulated market. An XGBoost machine learning surrogate model is trained to approximate the mapping from model parameters to stylised facts distance, and help to find the optimal parameters. Detailed implementations are presented below.

\subsubsection{Implementation}
\hspace{0.3cm} The whole calibration methodology proceeds with the following steps. 

\paragraph*{Step 1. Initialisation \\} 
\hspace*{0.3cm} The process starts with drawing a relatively large pool of parameter combinations. Each combination is a vector containing a value for each parameter: $\kappa$, $\beta$, and $\sigma_N$. The requirement is that the pool should be a good approximation of the whole parameter space. In our experiments, we use Sobol sampling (\citealp{morokoff1994quasi}) to implement the sampling routine. Sobol sampling is capable of guaranteeing uniformity of distribution even though the sampled set has a small number of points. It is shown that in terms of uniformity properties, Sobol sequences outperform the sequences generated by other sampling techniques such as Latin Hypercube sampling (\citealp{kucherenko2015exploring}). Other advantages of Sobol sampling include efficient implementation and faster sampling speed. The sampling number and quality of the pool of parameter combinations dominate the ability of the whole process to learn a good surrogate model. As a result, faster sampler is preferred so that more samples can be obtained in limited computational time. In our experiments, we use Sobol sampling to sample 16384 ($2^{14}$) points as the pool of parameter combinations.

\hspace{0.3cm} After the pool of parameter combinations is obtained, an initial set of samples is chosen randomly from the pool as the initial training set. Each point in the set of initialization samples is evaluated by actually running the agent-based model with the corresponding parameter combination. The corresponding stylised facts distance is calculated, which will act as the true label associated with that point in the training set. After this step we obtain an initial training set of parameter combinations with corresponding stylised facts distances as labels. The size of the initial training set in our experiments is 2000.

\paragraph*{Step 2. Surrogate Model Training \\} 
\hspace*{0.3cm} Given a training set of evaluated parameter combinations and corresponding stylised facts distances, an XGBoost regressor is learned over the training set in order to build the surrogate model. The input is the vector of model parameters to be calibrated, which in our experiments has dimension 3. The output is the stylised facts distance, which is a scalar. The XGBoost regressor is trained to fit the mapping from model parameters to stylised facts distance. Implemented under the Gradient Boosting framework, XGBoost is a machine learning algorithm designed to be highly flexible, efficient and portable (\citealp{chen2016xgboost}). The XGBoost algorithm builds an ensemble of simple decision trees, which are subsequently aggregated to improve the prediction performance. Details about the XGBoost algorithm can be found in \cite{chen2016xgboost}. 

\paragraph*{Remark 1.} One difficulty in training the XGBoost regressor is how to tune hyper-parameters of the XGBoost algorithm. Here we employ Bayesian Optimisation method based on Gaussian Process (\citealp{snoek2012practical}) to fine-tune the hyper-parameters of XGBoost. In the framework of Bayesian Optimization, performance of the XGBoost regressor is modelled as a sample from a Gaussian Process. The Gaussian Process then guides the exploration of the hyper-parameter space and helps to find an optimal set of hyper-parameters of XGBoost. Note that here the focus is the exploration of hyper-parameter space of the XGBoost machine learning algorithm, which is different from the parameter space of the agent-based model. Details about Bayesian Optimisation with Gaussian Process can be found in \cite{snoek2012practical}.

\paragraph*{Remark 2.} Another technique we have applied during the training process is to clip the values of training labels into a relatively small range. Specifically, for any training point, if the calculated stylised facts distance is too large, the distance value will be replaced with a relatively smaller value. The logic for this operation is that our calibration focus is the area of the model parameter space where stylised facts distances are small. In other words, parameter combinations with large stylised facts distances are of no interest and it is not important to have precise surrogate model approximations of stylised facts distances in those areas. If the range of distance value is not restricted, some large input labels would utterly bias the learning process of the XGBoost surrogate model. In this circumstances, the XGBoost surrogate model would wrongly try to fit those large outlier values. Consequently, the model is no longer a good approximation of the mapping from agent-based model parameters to stylised facts distance. By restricting the values of training labels to a relatively smaller range, the bias is successfully corrected and experimental results show that the surrogate model predicts the stylised facts distances quite precisely. Since optimal stylised facts distances in our experiments are generally smaller than 0.8, we restrict the training labels to the range of (0, 1].

\paragraph*{Step 3.  Surrogate Model Prediction\\} 
\hspace*{0.3cm} Once the surrogate XGBoost model is trained, it is used to predict the stylised facts distances over the set of remaining unlabelled parameter combinations. That is, the stylised facts distances that would be generated if those unlabelled parameter combinations were to be used in agent-based model simulation. The surrogate model predictions are used to guide further exploration of the model parameter space, as specified in the next step.

\paragraph*{Step 4. Supplement Training Set \\} 
\hspace*{0.3cm} Given the XGBoost surrogate model predictions, a subset of the unlabelled parameter combinations is drawn from the pool sampled in the first step. This subset of the unlabelled parameter combinations are evaluated in the agent-based model simulation, and the true stylised facts distances are calculated and subsequently assigned as the true labels of these samples. This set of newly labelled points is then added to the training set of labelled parameter combinations. There are two critical issues in this process: how many new points to be drawn into the subset of the unlabelled parameter combinations and how to select the points. For the first issue, the original method recommends that the number of new points to be drawn in this stage is the logarithm of the computational budget (\citealp{LAMPERTI2018366}). However, in our experiments it turns out that this rule would draw too few samples. Consequently, more iterations are required, which significantly reduces the computational efficiency. After lots of testing, in our method around 2 percent of total samples are drawn at each iteration, which in our case is 300 points. As for how to choose those unlabelled parameter combinations, we utilise the predictions made by the XGBoost surrogate model. The unlabelled points are sorted according to the predicted stylised facts distances. The points with smaller predicted stylised facts distances are selected, as the optimal parameter combination is more likely to exist among or near those points. However, not all points are selected according to predicted stylised facts distances. We also randomly choose some points from the unlabelled parameter combinations to avoid occasional bias induced by the XGBoost surrogate model. Specifically, 200 points are selected according to the "small predicted stylised facts distance" principle and 100 points are selected randomly.

\paragraph*{Remark 3.} The way of selecting the subset of unlabelled points implements an exploration-exploitation mechanism, which is a novel aspect of our methodology compared to the original method. Around two thirds of the points are selected according to the predicted stylised facts distances. In this way we exploit the information given by the XGBoost surrogate model and the model intelligently helps to direct the selection of new samples. The reason is that the optimal parameter combination is more likely to exist among the points with smaller predicted distances. This is true as long as the surrogate is a fairly good approximation of the real mapping from parameter combinations to stylised facts distances. However, exploration is also essential if the aim is to find a global minimum. There might be multiple local minimums inside the model parameter space. Completely exploiting the information given by the surrogate model may get the method stuck in a local minimum. Randomly sampling some unlabelled points helps to avoid this problem and contributes to better exploration of the whole model parameter space.

\paragraph*{Step 5. Iterations \\} 
\hspace*{0.3cm} After the training set is supplemented by the newly drawn samples and the corresponding labels, a new XGBoost surrogate model is trained using the new training set. The procedure is identical to previous steps. In other words, the previous "training - predicting - supplement"(step 2 to step 4) process is repeated until the budget of computational time is reached. In our experimental settings, we find that generally less than 5 iterations are required to build a fairly good surrogate model, whose stylised facts distance prediction error is less than 5\% at the predicted optimal point.

\section{Results and Evaluation}

\hspace{0.3cm} The whole methodology is run on 75 stocks from three exchanges: NASDAQ, LSEG, and HKEX. The main results are the stylised facts distances of the calibrated model, compared with the baseline stylised facts distances where model parameters are estimated by Expectation-Maximisation algorithm. We also present the error rate of XGBoost surrogate model prediction for stylised facts distance at the predicted optimal point to evaluate the surrogate model prediction accuracy. Finally, we show the methodology's capability of reproducing autocorrelation patterns in return series, which is an advantage of XGB-Chiarella methodology over other traditional models.

\subsection{Stylised Facts Distance}

\hspace{0.3cm} Table~\ref{actualdistancetable} shows the stylised facts distances of the calibrated model for all stocks, averaged by trading days. The corresponding standard deviations are also presented. For stocks in HKEX, the XGB-Chiarella method outperforms EM estimation algorithm for all 25 stocks. As for stocks in NASDAQ, in 22 out of 25 stocks the performance of XGB-Chiarella method is better than EM estimation method. On average, the XGB-Chiarella method achieves around 10\% smaller stylised distances than the EM baseline in this two exchanges, showing that the simulated market is calibrated to be more realistic. The standard deviations of the stylised facts distance are also generally a bit smaller for the proposed XGB-Chiarella method in NASDAQ and HKEX. For stocks in LSEG, performance of XGB-Chiarella method is not as good as the performance in the other two markets, with smaller advantage over the EM baseline estimation algorithm. However, it is still valid to say that the XGB-Chiarella methodology outperforms the baseline since smaller stylised facts distance is achieved in more than half of the stocks. Overall, the results show that the proposed XGB-Chiarella methodology outperforms the EM baseline in terms of stylised facts distance and the realism of the market simulation.

\hspace{0.3cm} Results in Table~\ref{actualdistancetable} are the mean and standard deviation of stylised facts distances across all trading days in our data span. To scrutinize each trading day for individual stock, Table~\ref{distancetable} shows the number of trading days when the XGB-Chiarella methodology outperforms the EM baseline in terms of stylised facts distance, and the percentage of these trading days out of all trading days. In Table~\ref{distancetable}, column "Better" represents the number of trading days when performance of XGB-Chiarella method is better than baseline EM estimation, while column "Total" represents the number of all trading days in our experiments. For the performance in different exchanges, results here are similar to the results in Table~\ref{actualdistancetable}. For most stocks in NASDAQ and HKEX, the XGB-Chiarella method performs better than the EM baseline for more than 85\% of total trading days. As for stocks listed in LSEG, this percentage is slightly lower, but is still on average more than 70\%. On the whole, for most trading days the XGB-Chiarella method is capable of creating a simulated financial market with smaller stylised facts distance, indicating that the proposed XGB-Chiarella method is able to generate more realistic artificial financial markets. 

\begin{table}[H]
\centering
\caption{Average stylised facts distance comparison between XGB-Chiarella calibration and Expectation-Maximisation method. The stylised facts distance is the average across all trading days for each stock; values in parentheses are corresponding standard deviations. There are several rare cases where distances are much larger, showing algorithms fail to obtain a reasonable set of parameters during the parameter estimation process.}
\vspace{0.3cm}
\label{actualdistancetable}
\csvreader[
  respect all,
  tabular=|m{1.2cm}<\centering|m{1.6cm}<\centering|m{1.7cm}<\centering|m{1.1cm}<\centering|m{1.7cm}<\centering|m{1.7cm}<\centering|m{1.1cm}<\centering|m{1.6cm}<\centering|m{1.6cm}<\centering|,
  table head=\hline NASDAQ & XGB-Chiarella & EM-algorithm & LSEG & XGB-Chiarella & EM-algorithm & HKEX & XGB-Chiarella & EM-algorithm  \\\hline,
  late after last line=\\\hline 
]{
  distance_actual.csv
}{}{\csvlinetotablerow}
\end{table}

\begin{table}[H]
\centering
\caption{Percentage of trading days when XGB-Chiarella calibration outperforms Expectation-Maximisation method.}
\vspace{0.3cm}
\label{distancetable}
\csvreader[
  respect all,
  tabular=|m{1.2cm}<\centering|m{0.7cm}<\centering|m{0.6cm}<\centering|m{1.4cm}<\centering|m{1.2cm}<\centering|m{0.7cm}<\centering|m{0.6cm}<\centering|m{1.4cm}<\centering|m{1.2cm}<\centering|m{0.7cm}<\centering|m{0.6cm}<\centering|m{1.4cm}<\centering|,
  table head=\hline NASDAQ & Better & Total & Percentage & LSEG & Better & Total & Percentage & HKEX & Better & Total & Percentage  \\\hline,
  late after last line=\\\hline 
]{
  distance.csv
}{}{\csvlinetotablerow}
\end{table}

\subsection{Surrogate Model Performance}

\hspace{0.3cm} To evaluate the performance of the surrogate model approximation, we present the prediction error rate when the surrogate model is used to predict the real stylised facts distance. Table~\ref{surrogatetable} shows a comparison between surrogate model predicted and actual stylised facts distance at the surrogate predicted optimal point in model parameter space for one trading day. Very similar results are achieved for other trading days. The results show that the XGBoost surrogate model is a very accurate proxy of the true model around the predicted optimal point for each stock on each trading day, with predicted stylised facts distance very close to actual stylised facts distance generated in agent-based model simulation. For most stocks, errors between predicted distance and actual distance are less than 10\%. It is also shown that the accuracy of XGBoost surrogate model prediction is similar across the three exchanges, which indicates the robustness of the XGB-Chiarella methodology. 

\begin{figure}[H]
\centering
\includegraphics[width=16.5cm, height=5.2cm, angle=0]{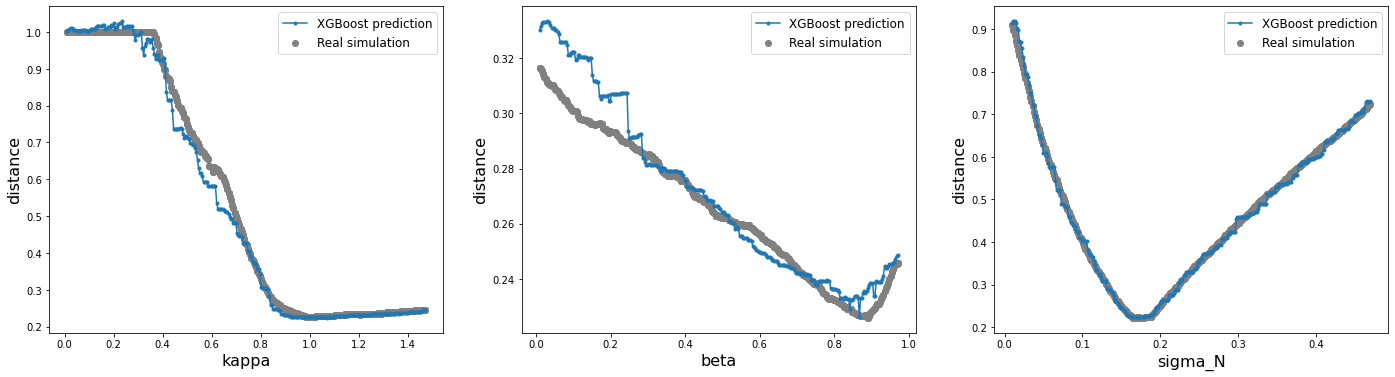}
\caption{Sensitivity analysis around the optimal point predicted by the XGBoost surrogate model. This surrogate model corresponds to the trading day on the 29th of September, 2021 for stock FB. For each sub-graph, model parameters are fixed to the optimal value except for the parameter in x-axis. Gray lines represent the actual stylised facts distances while blue lines represent the XGBoost surrogate model prediction for the stylised facts distances.}
\label{sensitivitygraph}
\end{figure}

\hspace{0.3cm} Apart from the prediction accuracy at the predicted optimal point, we also examine the prediction accuracy around the optimal point as a sensitivity analysis. Figure~\ref{sensitivitygraph} shows the comparison between XGBoost surrogate model prediction and actual simulated stylised facts value. For each sub-graph, one model parameter is assigned a series of values across the given range during calibration. Other model parameters are fixed to the optimal value. In this way a set of parameter combinations is obtained, with different values for that particular model parameter. Then predicted and actual stylised facts distance are calculated and compared, as shown in Figure~\ref{sensitivitygraph}. The gray lines are the actual stylised facts distances while blue lines represent the XGBoost surrogate model prediction for the stylised facts distances. It is shown that the XGBoost prediction line fits the actual simulated stylised facts distance line quite well, especially for parameter "sigma\_N". For each parameter, there is a unique minimal value in the graph, which corresponds to the optimal point predicted by the surrogate model. The results here indicate that the surrogate model approximates the real stylised facts distance quite accurately and is capable of finding the true optimal model parameter combination that is able to generate realistic financial market simulations.

\begin{table}[H]
\centering
\caption{Comparison between surrogate model predicted and actual stylised facts distance at the surrogate predicted optimal point. The results shown here are for one trading day (2nd of November, 2021), while results for other trading days are very similar.}
\vspace{0.3cm}
\label{surrogatetable}
\csvreader[
  respect all,
  tabular=|m{1.2cm}<\centering|m{0.8cm}<\centering|m{0.8cm}<\centering|m{1.1cm}<\centering|m{1.2cm}<\centering|m{0.8cm}<\centering|m{0.8cm}<\centering|m{1.1cm}<\centering|m{1.2cm}<\centering|m{0.8cm}<\centering|m{0.8cm}<\centering|m{1.1cm}<\centering|,
  table head=\hline NASDAQ & Predict & Actual & Error & LSEG & Predict & Actual & Error & HKEX & Predict & Actual & Error  \\\hline,
  late after last line=\\\hline 
]{
  surrogate.csv
}{}{\csvlinetotablerow}
\end{table}

\begin{figure}[H]
  \begin{subfigure}{0.33\textwidth}
    \includegraphics[width=5.5cm, height=4cm]{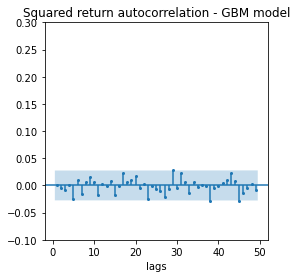}
    \caption{GBM}
  \end{subfigure}%
  \hspace*{\fill}   
  \begin{subfigure}{0.33\textwidth}
    \includegraphics[width=5.5cm, height=4cm]{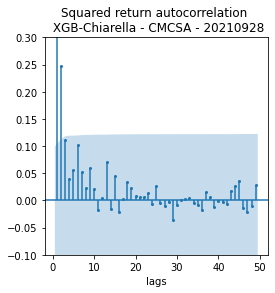}
    \caption{XGB-Chiarella}
  \end{subfigure}%
  \hspace*{\fill}   
  \begin{subfigure}{0.33\textwidth}
    \includegraphics[width=5.5cm, height=4cm]{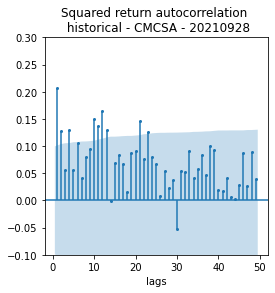}
    \caption{Historical}
  \end{subfigure}

\caption{\textbf{(a)}: Autocorrelation of squared returns generated by a GBM. \textbf{(b)}: Autocorrelation of squared returns generated by XGB-Chiarella method for CMCSA on Sept 28th, 2021. \textbf{(c)}: Historical autocorrelation of squared returns for CMCSA on Sept 28th, 2021.} 
\label{acf2comparison}
\end{figure}

\subsection{Autocorrelations}

\hspace{0.3cm} One unique advantage of agent-based financial market simulation is the capability of reproducing autocorrelation patterns of returns. Currently, most literature in financial econometrics model financial price series as Geometric Brownian Motion (GBM). For example, the Black-Scholes model for option pricing explicitly assumes the stock price follows a GBM. However, the case is not true in real financial market. One obvious anomaly is the autocorrelation patterns in return series, such as the volatility clustering phenomenon shown in Figure~\ref{acf2graph}. Our XGB-Chiarella method has an advantage over GBM model in that it can successfully reproduce the autocorrelation patterns in financial return time series. Figure~\ref{acf2comparison} shows autocorrelations of squared returns of a GBM model and that of a simulation under XGB-Chiarella methodology for one stock. The real historical autocorrelation of squared returns  on the same day for that stock is also presented. It is obvious that our XGB-Chiarella method outperforms the traditional GBM model in terms of the replication of realistic autocorrelation patterns, especially for small time lags. For small time lags, significant positive autocorrelations of squared returns exist in historical financial market data. This stylised fact is successfully reproduced in XGB-Chiarella simulation, while the autocorrelations of squared returns generated by a GBM is basically close to zero regardless of time lags. In fact, there is no specific patterns in autocorrelations generated by GBM model. Notice that simulated results in other stocks for other trading days are similar to what is shown here.

\begin{figure}[H]
\centering
\includegraphics[width=15cm, height=5cm, angle=0]{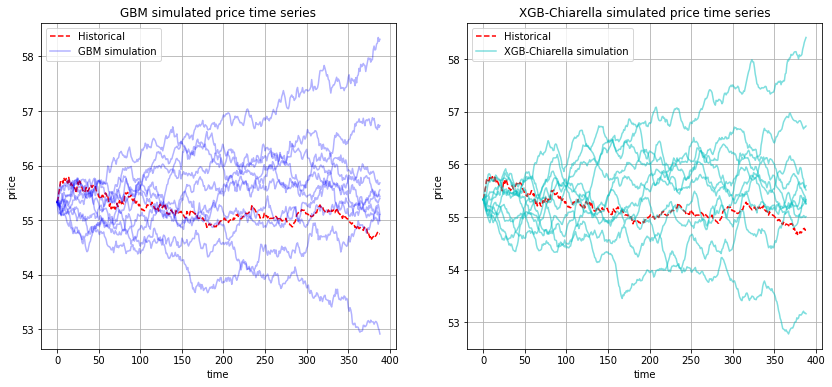}
\caption{Simulated price scenarios generated by GBM and XGB-Chiarella, respectively.}
\label{pricecompare}
\end{figure}

\begin{figure}[H]
\centering
\includegraphics[width=10cm, height=5cm, angle=0]{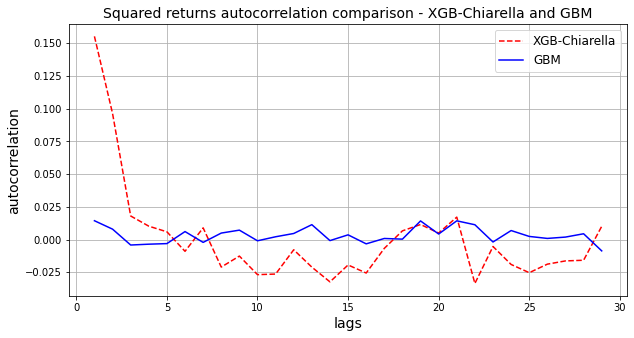}
\caption{Average autocorrelation of squared returns associated with the simulated price scenarios in Figure~\ref{pricecompare}. Blue line represents the GBM simulated scenarios while red dashed line represents the XGB-Chiarella simulated scenarios.}
\label{meanacfcompare}
\end{figure}

\hspace{0.3cm} Successful reproduction of autocorrelation patterns gives the XGB-Chiarella method an advantage in realistic simulation of multiple price series, which can be used in scenario simulations for risk management. When calculating simulation-based risk metrics such as value at risk (VaR), lots of institutions are still using GBM to simulate future scenarios. As shown in Figure~\ref{acf2comparison}, the return series simulated by GBM model lack realistic autocorrelation pattern, which would undermine the credibility of the risk metric calculation. Our proposed method is a better alternative. The proposed XGB-Chiarella method extracts fundamental values and calibrate the model parameters using historical data. With the same set of calibrated model parameters, the agent-based model is capable of generating multiple different scenarios by changing the input fundamental value series. For example, if GBM scenarios are used as fundamental value, the agent-based model is able to reproduce similar price series as the GBM model, but with realistic autocorrelation patterns of returns. Figure~\ref{pricecompare} shows the price scenarios generated by GBM and XGB-Chiarella method, respectively. Here the fundamental values in XGB-Chiarella model are the price scenarios generated by the GBM. Figure~\ref{meanacfcompare} shows the corresponding average autocorrelation of squared returns for the two cases. It is shown that the two models can generate similar price scenarios, but the XGB-Chiarella model is able to reproduce much more realistic autocorrelation patterns. Specifically, for small time lags the XGB-Chiarella model reproduces significant positive autocorrelation of squared returns, which is consistent with empirical data. In contrast, there are no significant autocorrelation patterns for GBM scenarios. It is believed that such agent-based price series simulation could provide a richer environment than GBM model for risk management practice such as VaR calculation. Since the focus of this paper is the intra-day price formation process, we will address this topic in a separate paper in the future. 

\section{Conclusion and Future Work}
\subsection{Summary of Achievements}
\hspace{0.3cm} In this paper, a new approach called XGB-Chiarella is proposed to generate realistic intra-day artificial financial price data in order to provide insight into the intra-day price formation process. To the best of our knowledge, this is the first extension of the Chiarella model to generate minute-level intra-day financial market simulation. The approach utilises agent-based modelling techniques. The underlying simulation model has only three agents: one for fundamental trader, one for momentum trader, and one for noise trader. The model is simulated and model parameters are calibrated by an XGBoost machine learning surrogate. The proposed methodology is tested on 75 stocks from three exchanges: NASDAQ, LSEG, and HKEX. In terms of stylised facts distance, the proposed XGB-Chiarella method is able to generate more realistic financial market simulations than the original Expectation-Maximisation estimation algorithm. This is true in nearly all the stocks from three exchanges. Despite the fact that the methodology is based on a model with only three agents, the XGB-Chiarella methodology successfully generates very realistic financial market simulations. This indicates that one agent per category seems to be sufficient to capture the intra-day price formation process for the time scale (minutes) chosen in this paper. The very simple model structure not only accelerates the simulation process in terms of computational cost, but also enable us to scrutinize the intra-day price formation process, such as the trend and value effects. The results provide support for the existence of a universal intra-day price formation mechanism. The realistic simulated intra-day financial market indicates that trend and value effects, as well as noise trading, are indispensable to the intra-day price formation process.

\hspace{0.3cm} We also show that in the process of calibration, the XGBoost surrogate model is an accurate approximation of the true model. At the predicted optimal point for each stock on each trading day, the surrogate model prediction error is mostly smaller than 10\%. The machine learning surrogate is capable of intelligently directing the exploration of model parameter space. The exploitation-exploration mechanism is also introduced in the model calibration process. A practical application of the proposed methodology is also presented.

\subsection{Future Work}
\hspace{0.3cm} This work can be extended in several aspects. Firstly, in modern financial markets a large number of transactions can happen in fractions of a second, raising interest in the price formation process at higher frequency. Therefore it would be interesting to test whether the proposed methodology would work at higher frequency, for example in microseconds or even nanoseconds level. Another interesting extension is about agent heterogeneity. For example, the momentum traders can be divided into two groups: one group of traders that focus on lower frequency price momentum and the other group that acts according to the value of higher frequency price momentum. It is expected that the introduction of further agent heterogeneity would improve the realism of the model since real-world traders are obviously heterogeneous. In addition, it would also be interesting to understand the price behaviours if we relax the assumption of $\lambda$-approximation in the underlying Chiarella model. With the $\lambda$-approximation, it is assumed that price change is linearly proportional to the cumulative demand of all traders. One extension to relax this assumption is to introduce full exchange protocols and limit order books to the simulated financial market and investigate the corresponding market dynamics. In this circumstance, hundreds of agents can be included in the model and interact with each other through limit order books, which is exactly the mechanism existing in real financial markets. The last aspects of future work involves extending the Chiarella model to multiple stocks. For example, how to simulate the correlation structure across multiple stocks and how to create correlated demands for related stocks during simulation.


\section*{Acknowledgements}
\hspace{0.3CM} The support of the UK EPSRC (Grant Nos. EP/L016796/1, EP/N031768/1, EP/P010040/1,  EP/S030069/1 and EP/V028251/1), Xilinx, and Intel is gratefully acknowledged. Kang Gao holds a China Scholarship Council-Imperial Scholarship. The authors thank Kaveh Aasaraai and Javier A. Varela for their excellent suggestions. 

\bibliographystyle{myapalike}
\bibliography{references}



\end{document}